\documentclass[onecolumn,draftclsnofoot]{IEEEtran}
\usepackage{times,amsmath,epsfig,latexsym,amssymb,psfrag,booktabs,extpfeil,graphicx,subfig,hyperref,setspace}
\begin{document}
 
\doublespacing

\title{Performance Analysis of Ad-Hoc Routing in Clustered Multi-hop Wireless Networks}
\author{Amin Azari$^*$, Jalil S. Harsini$^+$ and Farshad Lahouti$^*$(Corresponding Author)\\
$^*$Center for Wireless Multimedia Communications,\\ School of Electrical \& Computer Engineering, University of Tehran\\
$^+$Department of Electrical Engineering, University of Guilan\\
Email: \{aazari,lahouti\}@ut.ac.ir, harsini@guilan.ac.ir, WEB: http://wmc.ut.ac.ir}
\maketitle

\begin{abstract}
This paper analyzes the performance of clustered decode-and-forward multi-hop relaying (CDFMR) wireless Rayleigh fading networks, and sheds light on their design principles for energy and spectral efficiency. The focus is on a general performance analysis (over all SNR range) of heterogeneous wireless networks with possibly different numbers of relays in clusters of various separations. For clustered multi-hop relaying systems, ad-hoc routing is known as an efficient decentralized routing algorithm which selects the best relay node on a hop-by-hop basis using local channel state information. In this article, we combine ad-hoc routing and cooperative diversity in CDFMR systems, and we derive (i) a closed-form expression for the probability distribution of the end-to-end SNR at the destination node; (ii) the system symbol error rate (SER) performance for a wide class of modulation schemes; and (iii) exact analytical expressions for the system ergodic capacity, the outage probability and the achievable probability of the SNR (power) gain. We also provide simple analytical asymptotic expressions for SER and the outage probability in high SNR regime. Simulation results are provided to validate the correctness of the presented analyses.
\end{abstract}

\IEEEpeerreviewmaketitle
\section{Introduction}
\newcommand{\ud}{\mathrm{d}}

\IEEEPARstart{W}{ireless} networks with multi-hop transmission
capability facilitate forwarding packets from one node to other
nodes in the network that may not be within direct wireless
transmission range of each other. Recently, different
aspects of multi-hop wireless networks, including both mobile ad-hoc
networks (MANETs) and hybrid cellular networks, are explored from 
signal processing, networking and information theory perspectives and prototypes are developed (see, e.g.,  \cite{ad3}-\nocite{ad2}\cite{ad1} and the references therein). 
Among three main relaying schemes, i.e., decode-and-forward (DF), compress-and-forward, and
amplify-and-forward (AF), the DF protocol has the benefit of avoiding
noise propagation, and therefore it provides a better link
reliability. In addition, when in multi-hop transmission
systems the cooperative relay in each hop may be selected among
several adjacent relays, in a so-called relay cluster, better transmission
reliability may be provided by achieving diversity gain through the
cooperation among cluster nodes. In this scheme, each relay node
participates in an ad-hoc routing protocol that allows to discover
the best multi-hop path through the network towards the destination node.
In practice, the networks are heterogeneous in the sense that they are composed of multiple clusters with possibly different distances and numbers of relays.
This paper analyzes the performance of
heterogeneous clustered decode-and-forward multi-hop relaying (CDFMR) wireless networks and aims to quantify how design parameters affect the network energy and spectrum efficiencies.
\subsection{Previous Works}
There have been significant efforts on the study of cooperative
systems, e.g., \cite{kram}-\nocite{lan}\cite{po3}, and more also on combining routing and cooperative
diversity in multi-hop fading wireless networks \cite{po4} \nocite{po7}\nocite{po8}\cite{po9}. In \cite{ble}\cite{jad}, for
dual-hop cooperative systems, composed of one source-destination
pair and a number of relays, the problem of finding the transmission path that achieves full diversity and the
associated performance analysis are studied under AF and DF
relaying protocols. Relay protocol design for IEEE 802.16 relay networks is considered in \cite{po1}-\cite{po2}. Deplying relays in cellular network is considered in \cite{po5}-\cite{po6} for satisfying users’ increased bit rate requirements while still retaining the benefits of a cellular structure.
Addressing the same problem in general multi-hop networks with more than
two hops is of great interest and is studied in, e.g., \cite{del}\cite{yon}. 
The main idea is to group the relay nodes in each hop into clusters and develop multi-hop transmission
protocols based on data communication between adjacent
cluster heads. 
Following this approach, several routing
strategies and their asymptotic outage probability performance in
high SNR regime are investigated for CDFMR systems in \cite{del}.
As shown in \cite{del}, in a DF multi-hop network with $L$ cooperating relays
per hop, the maximum diversity order is $L$ regardless of the number of hops. 
This is where the optimal routing strategy is defined as the one which identifies the path with the
minimum end-to-end outage probability among all possible paths. In fact, the resulting routing protocol aims at finding the path with the greatest minimum SNR among hops, while relying on the knowledge of the channel state information (CSI) of all links. 
In \cite{den}, it is shown that arbitrary relay selection in the
first $N-1$ clusters of an $N$ cluster CDFMR system, and relay selection based on channel quality in the final cluster could achieve the same diversity gain as the optimal routing algorithm. Although, this leads to much lower power gain, but it only relies on the CSI of the last hop.  
A decentralized and more efficient
ad-hoc routing strategy is also reported in which the relay
selection is performed in a per-hop manner using only the CSI of
relay links in the associated cluster. As verified in
\cite{del}, when the number of hops in the network is small, the ad-hoc routing
provides an outage performance very close to that of the
optimal routing strategy, making it an attractive protocol with low
algorithmic complexity and overhead.

The problem of maximizing the achievable rates by selecting a relaying
subset and the allocation of transmission time in DF multi-relay
systems is investigated in \cite{ber}. In \cite{bab},
efficient routing algorithms for linear multi-hop networks are presented, which minimize the end-to-end
system outage probability when equal-power or optimal-power
allocation at the physical layer is employed. The performance
analysis for an interference-aware opportunistic relay selection
protocol in a multi-hop line network
is considered in \cite{sta}. 

As relay assisted transmission has shown its merits for data transfer
purposes, most recent wireless standards already provision for multi-hop relay networking \cite{802}- \nocite{lte1}\cite{jiaicc}. It is expected that multi-hop networks also play key roles in next generation wireless systems for example in backhauling wireless mesh networks.

\subsection{Outline of Contributions and Structure of the Article}
In this work, we present an exact analysis of the performance of ad-hoc
routing over heterogeneous clustered decode-and-forward multi-hop relaying wireless networks. The proposed analyses quantify the performance in terms of outage probability, ergodic capacity
and symbol error rate (SER) over all range of channel conditions. 
We combine ad-hoc routing and cooperative diversity in CDFMR systems, with the consideration of a
more realistic system and channel model. In particular, in the
considered system model the number of relays in each cluster in
addition to the distance between the relay clusters may be
different, making the model more general compared to the prior art. Considering this
general CDFMR network model, the main contributions of the current article are as follows:

\begin{enumerate}
\item
We derive a closed-form expression for the probability distribution function (PDF) of the end-to-end
SNR at the destination node.
\item
Using the PDF in item 1, we derive exact analytical
expressions for the system outage probability, the ergodic capacity,
and the achievable probability of the SNR gain. Furthermore, we
analyze the system symbol error rate performance for a wide class of modulation schemes. The first two metrics quantify how efficiently the spectrum is utilized and the last two are indicators of energy efficiency of the multi-hop network.
\item
We also investigate the asymptotic behavior of the outage
performance and SER in high SNR regime. In this line, we present
simple analytical asymptotic expressions for SER and the outage
probability.
\end{enumerate}
Moreover, simulation results are provided to validate the
correctness of the proposed analytical performance assessments.

The rest of this paper is organized as follows. In the next Section, the system model is introduced. The exact performance of CDFMR network with ad-hoc routing in terms of outage, SER, rate, and probability of SNR gains is presented in Section III. In Section IV, the asymptotic performance analyses are presented. Section V is devoted to numerical results and discussions. The concluding remarks are given in Section VI.

\begin{figure}[!t]
\centering
\includegraphics[width=3.5in]{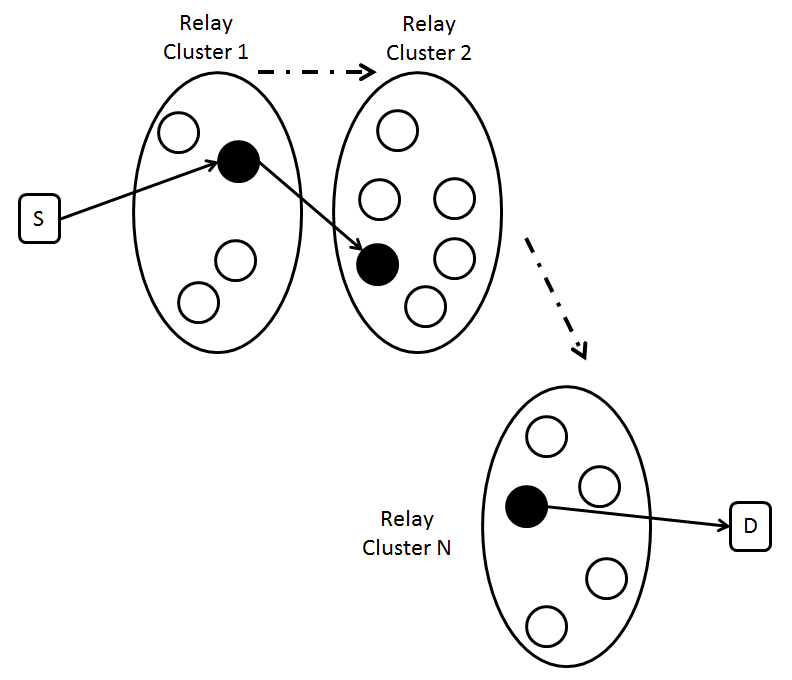}
\caption{System model. Ad-Hoc routing in CDFMR network is considered
(see Section II for details).} \label{relsel}
\end{figure}

\section {System Model}\label{SystemModelSection}
As shown in Fig. 1, we consider a $N+1$ hop clustered network
model in which $N$ intermediate relay clusters are placed on the
path between the source and the destination nodes. In this model,
the $i$th relay cluster includes $L_i$ single antenna relay nodes,
out of which only one node is selected during each time-slot
interval. To model the wireless link between any two nodes in the
system we adopt a narrowband Rayleigh fading channel model with
additive white Gaussian noise (AWGN). The fading gains of different
relay links in each hop are assumed to be statistically independent,
as the relays are usually spatially well separated. Moreover, we
assume that the distance between clusters is much larger than those
between the relay nodes in each cluster, hence the links in each hop
are modeled with the same Rayleigh distribution, which is
independent from the other hops. The objective of the system is to
transmit a data packet generated at the source node to the
destination node. To this end, a time division multiple access (TDMA) scheme is adopted so that at
each time-slot only transmission in one hop is active and relays
have the half duplex constraint. The relays are constant power and the fading channel gains are
assumed constant during one time-slot duration.

In the ad-hoc routing algorithm, the relay selection is performed in
a per-hop manner, thus at the $i$th hop only the CSI of $L_i$ links is
needed \cite{del}. In each of the first $N-1$ hops, only the best
relay, i.e, the one with the highest received SNR, is selected to
forward the source packet. The routing in the last two hops is
different, as there is only one destination node (D) in the last
hop. In fact, to achieve the maximum diversity gain relay selection
in the $N$th hop takes into account the CSI of the links in both $N$
and $N+1$ hops. Table \ref{aLg} summarizes the ad-hoc routing
algorithm in the CDFMR network based on the following notations: The
capital letter S refers to the source node, $r^*_i$ indicates the
index of the selected relay node in the $i$th hop, and
$\gamma_{r^*_{i-1},j,i}$ is the received SNR for the link from the
selected node in the $(i-1)$th hop to the $j$th node in the $i$th
hop.

\begin{table}[!t]
\centering \caption{Ad-Hoc routing algorithm in CDFMR network. The
relays selected in different hops $\{r^*_1,\hdots,r^*_N\}$ set the
route from the source to the destination.}\label{aLg}
\begin{tabular}{p{0.2 cm}p{7.5 cm}}
\toprule[0.5mm]
1. &$ r^*_0 =S$\\
2. & for $i=1:N-1$ \\
& $r^*_i=\arg \max_{l=1,\hdots,L_i}(\gamma_{r^*_{i-1},l,i})$\\
& end\\
3. & $r^*_N=\arg \max_{l=1,\hdots,L_N}\min (\gamma_{r^*_{N-1},l,N}, \gamma_{l,D,N+1})$\\
\bottomrule[0.5mm]
\end{tabular}
\end{table}

\section{Exact Performance Analysis}
In this Section, we first derive the PDF of the end-to-end SNR of
the CDFMR system employing the ad-hoc routing protocol. We then use
this PDF to derive closed-form expressions for the outage
probability, the ergodic capacity and the SER performance of the
considered system model.
\subsection{PDF of the End-to-End SNR}\label{PDF}
For the purpose of performance evaluation, we need the PDF of the
received SNR at the destination node. For DF multi-hop relaying
transmission, the instantaneous data rate is the minimum rate
between the $N$+1 hops \cite{jadid}. As a result, the hop with the
lowest SNR creates a performance bottleneck. Hence, the equivalent
end-to-end received SNR is expressed as
\begin{equation}
\gamma_t = \min_{i=1:N} (\gamma^i_t)
\end{equation}
where based on the relay selection rule in Table \ref{aLg}, we
define
\begin{equation} \gamma^i_t =
\max_{l=1,\hdots,L_i}(\gamma_{r^*_{i-1},l,i}) \hspace{4mm}
\text{for}\hspace{2mm} i=1:N-1
\end{equation}
and $\gamma^N_t$ is the received SNR over the last two hops,
expressed as
\begin{equation}
\gamma^N_t =\max_{l=1,\hdots,L_N}\min (\gamma_{r^*_{N-1},l,N}, \gamma_{l,D,N+1}).
\end{equation}
Using (1), the cumulative density function (CDF) of the variable
$\gamma_t$ can be written as
\begin{align}\label{e1}
F_{\gamma_t}(x)&=\Pr(\gamma_t \le x) = \Pr(\min_{i=1:N}(\gamma^i_t) \le x)\nonumber\\
&=1-\Pr(\min_{i=1:N}(\gamma^i_t) \ge x).
\end{align}
Since  SNRs of different links are independent random variables, one can
rewrite \eqref{e1} as
\begin{equation}\label{e2}
F_{\gamma_t}(x)=1-\prod\limits_{i = 1}^N \Pr({\gamma^i_t \ge x}).
\end{equation}
\begin{equation}\label{e2.1}
= 1 - (1 - {F_{\gamma _t^1}}(x))(1 - {F_{\gamma _t^2}}(x)) \cdots (1
- {F_{\gamma _t^{N - 1}}}(x))(1 - {F_{\gamma _t^N}}(x))
\end{equation}
Since the fading gain of each channel at the $i$th hop has Rayleigh
distribution, the CDF of the SNR $\gamma_{r^*_{i-1},l,i}$
is given by
$F_{\gamma_{r^*_{i-1},l,i}}(x)=1-\exp\left(-\frac{x}{\Gamma_i}\right)$,
where $\Gamma_i$ is the expected value of all link SNRs at the $i$th
hop. Using order statistics \cite{order}, and the binomial
expansion, the CDF $F_{\gamma^i_t} (x)$ in \eqref{e2.1} can then be
obtained as follows
\begin{equation}\label{e5}
\begin{array}{l}
F_{\gamma^i_t}(x) = \left(1-\exp\left(-\frac{x}{\Gamma_i}\right)\right)^{L_i}\\
\hspace{11mm}=\sum\limits_{j = 0}^{L_i} {\left(_j^{L} \right)}(-1)^j
\exp\left(-j\frac{x}{\Gamma_i}\right),
\hspace{5mm}\text{for}\hspace{3mm} i=1,...,N-1.
\end{array}
\end{equation}
Also, using (3) the CDF $F_{\gamma^N_t}(x)$ is obtained as
\begin{equation}\label{e4}
\begin{array}{l}
F_{\gamma^N_t}(x)=[1-(1-F_{\gamma_{N}}(x))(1-F_{\gamma_{N+1}}(x))]^{L_N}\\
\hspace{12mm}{=}\sum\limits_{j = 0}^{L_N} {\left(
_j^{L_N} \right)}(-1)^j \exp\left(-j\frac{x}{\Gamma_N}-j\frac{x}{\Gamma_{N+1}}\right)\\
\hspace{12mm}{=}\sum\limits_{j = 0}^{L_N} {\left( _j^{L_N}
\right)}(-1)^j \exp\left(-j\frac{x}{\Gamma_N^t}\right)
\end{array}
\end{equation}
where we define
$\Gamma^t_N=\frac{\Gamma_N\times\Gamma_{N+1}}{\Gamma_N+\Gamma_{N+1}}$, and 
$\Gamma^t_i=\Gamma_i$ for $i=1:N-1$, as will be used in the sequel. 
Substituting \eqref{e5} and \eqref{e4} into \eqref{e2.1} yields
\begin{equation}\label{e6}
\begin{array}{l}
{F_{{\gamma _t}}}(x) = 1 - \prod \limits_{i=1}^{N}\left[1-\sum\limits_{j = 0}^{L_i} {\left(_j^{L_i} \right)}(-1)^j \exp\left(-j\frac{x}{\Gamma^t_i}\right)\right].
\end{array}
\end{equation}
Using the binomial expansion and following some mathematical
manipulations, the CDF in \eqref{e6} can be rewritten as
\begin{equation}\label{tw-span}
\begin{array}{l}
{F_{{\gamma _t}}}(x) 
 =1 - (-1)^N \prod \limits_{i=1}^{N}\left[(\sum\limits_{j = 1}^{L_i} {\left(_j^{L_i} \right)}(-1)^j \exp\left(-j\frac{x}{\Gamma^t_i}\right))\right]\\
\hspace{11mm}=1 +\sum\limits_{{j_1} = 1}^{L_1} {\sum\limits_{{j_2} = 1}^{L_2} {
\cdots \sum\limits_{{j_N} = 1}^{L_N}
{\left(_{j_1}^{L_1}\right)\left(_{j_2}^{L_2}\right) \cdots
\left(_{j_N}^{L_N}\right)( - 1)^{( {j_1} +  \cdots  + {j_N}+N+1)}
\exp \left( - Kx\right)} } }
\end{array}
\end{equation}
where $K=\sum \limits_{i=1}^{N} \frac{j_i}{\Gamma^t_i}$.\\
The PDF of the end-to-end received SNR is obtained by computing the
derivative of ${F_{{\gamma _t}}}(x)$ with respect to $x$, as follows
\begin{equation}\label{PdF}
\begin{array}{l}
{f_{{\gamma _t}}}(x) =\sum\limits_{{j_1} = 1}^{L_1} \sum\limits_{{j_2} = 1}^{L_2}  \cdots \sum\limits_{{j_N} = 1}^{L_N} \left(_{j_1}^{L_1}\right)\left(_{j_2}^{L_2}\right) \cdots \left(_{j_N}^{L_N}\right)( - 1)^{( {j_1} +  \cdots  + {j_N}+N)} K{\exp \left( - Kx\right)}  .
\end{array}
\end{equation}

\subsection{Ergodic Capacity}\label{ergodic}
Using the derived PDF in \eqref{PdF}, the ergodic capacity of the
considered CDFMR system can be computed in bits/s/Hz as follows
\begin{equation}\label{cerg}
\begin{array}{l}
{C_E} = \frac{1}{{N + 1}}\int_0^\infty {\log_2 (1 + x) {f_{{\gamma _t}}}(x) dx }\\
\hspace{5.5mm}= \frac{1}{{(N + 1)\ln2}}\int_0^\infty {\ln (1 + x) {f_{{\gamma _t}}}(x) dx}\\
\hspace{5.5mm}=\frac{1}{{(N + 1)\ln2}}\sum\limits_{{j_1} = 1}^{L_1} \sum\limits_{{j_2} = 1}^{L_2}  \cdots \sum\limits_{{j_N} = 1}^{L_N} \left(_{j_1}^{L_1}\right)\left(_{j_2}^{L_2}\right) \cdots \left(_{j_N}^{L_N}\right){( - 1)^{( {j_1} +  \cdots  + {j_N}+N)}}K\int_0^\infty {\ln (1 + x)\exp \left( - Kx\right)dx}  ,
\end{array}
\end{equation}
where the factor $\frac{1}{N+1}$ is due to the half duplex operation
of relays. Using the equality $\int_0^\infty {\ln (1 +
x)\exp\left(-ax\right)dx =} \frac{{\exp (a){\text{E}_1}(a)}}{a}$,
the ergodic capacity can be expressed as follows
\begin{equation}\label{eRg}
\begin{array}{l}
{C_E} = \frac{1}{{(N + 1)\ln2}}\sum\limits_{{j_1} = 1}^{L_1} \sum\limits_{{j_2} = 1}^{L_2}  \cdots \sum\limits_{{j_N} = 1}^{L_N} \left(_{j_1}^{L_1}\right)\left(_{j_2}^{L_2}\right) \cdots \left(_{j_N}^{L_N}\right){( - 1)^{( {j_1} +  \cdots  + {j_N}+N)}} \exp (K){\text{E}_1}\left(K\right)
\end{array}
\end{equation}
where ${{\rm{E}}_1}(x) = \int_x^\infty  {\frac{{{e^{ - t}}}}{t}} dt$
is the exponential integral \cite{tab2007}, and $K$ is defined in
Subsection III.A.

\subsection{Outage Probability}\label{outage}
For the CDFMR network under consideration, we define the outage
probability as the probability that the instantaneous end-to-end
transmission rate falls below a predefined target rate, denoted by
$R_{th}$. The instantaneous end-to-end rate can be written as
\begin{equation}\label{eq16}
\begin{array}{l}
R = \frac{1}{{N + 1}}\log (1 + {\gamma _t}).
\end{array}
\end{equation}
Hence, the outage probability is given by
\begin{equation}\label{eq17}
\begin{array}{l}
P_{out} = \Pr(R < R_{th})\\
\hspace{7.5mm}=\Pr\left(\gamma_t < 2^{(N + 1)R_{th}} - 1\right).
\end{array}
\end{equation}
Let $A = {2^{(N + 1)R_{th}}} - 1$. Using the CDF of the SNR $\gamma_t$ in (10), the probability in \eqref{eq17} is obtained
as
\begin{equation}\label{oUt}
\begin{array}{l}
 P_{out}  = 1 +\sum\limits_{{j_1} = 1}^{L_1} {\sum\limits_{{j_2} = 1}^{L_2} { \cdots \sum\limits_{{j_N} = 1}^{L_N} {\left(_{j_1}^{L_1}\right)\left(_{j_2}^{L_2}\right) \cdots \left(_{j_N}^{L_N}\right)( - 1)^{( {j_1} +  \cdots  + {j_N}+N+1)} \exp \left( - A K\right)} } }

\end{array}
\end{equation}

It is noteworthy that the outage probability expression in
\eqref{oUt} is exact and novel for the CDFMR network with arbitrary
number of hops. For $N$=1, i.e., dual-hop transmission system, the
outage probability expression in \eqref{oUt} reduces to (5) in
\cite{ado}.

\subsection{SER Performance}\label{SER}
In this Section, we derive the symbol error probability for the
CDFMR network with ad-hoc routing and different modulation schemes. In general, the SER
performance for a wide class of modulation schemes can be calculated
as \cite{dig}
\begin{equation}\label{MOD}
\begin{array}{l}
SER =2 \alpha \int_0^\infty  {Q(\sqrt {2\beta x} ){f_{{\gamma _t}}}(x)dx} \\
\\
\hspace{9mm} = {\alpha }\int_0^\infty  {{\rm{erfc}}\left( {\sqrt {\beta x} } \right){f_{{\gamma _t}}}(x)dx}
\end{array}
\end{equation}
where $Q(\cdot)$ is the $Q$ function, $\text{erfc}(\cdot)$ is the
complementary error function, and the parameters $\alpha$ and
$\beta$ depend on the specific modulation scheme being used. Table
II gives these parameters for some common modulation schemes
\cite{dig}\cite{m2012}.
\begin{table}[!t]
\centering \caption{Parameters for SER in \eqref{MOD}}\label{CoefTable}
\begin{tabular}{p{2 cm}p{1.7 cm} p{1.5 cm}p{1.5cm}}\\
\toprule[0.5mm]
\textbf{Modulation} &\textbf{$\alpha$} &\textbf{$\beta$}&\textbf{Comment}\\
\midrule
BPSK &0.5 &1&Exact \\
MPSK, $M>$2 &1 &${\sin ^2}(\frac{\pi }{M})$&Approx. \\
MQAM &$2(1-\frac{1}{\sqrt{M}})$ &$\frac{3}{2(M-1)}$&Approx. \\
BFSK &0.5 &0.5&Exact \\
MPAM &$\frac{M-1}{M}$ &$\frac{3}{M^2-1}$&Exact \\
\bottomrule[0.5mm]
\end{tabular}
\end{table}
Substituting the end-to-end channel PDF of CDFMR network in
\eqref{PdF} into \eqref{MOD} leads to
\begin{equation}
\begin{array}{l}
SER =\alpha\sum\limits_{{j_1} = 1}^{L_1} \sum\limits_{{j_2} = 1}^{L_2}  \cdots \sum\limits_{{j_N} = 1}^{L_N} \left(_{j_1}^{L_1}\right)\left(_{j_2}^{L_2}\right) \cdots \left(_{j_N}^{L_N}\right)( - 1)^{( {j_1} +  \cdots  + {j_N}+N)} K
 \int_0^\infty \text{erfc}\left(\sqrt {\beta x} \right){\exp \left( - Kx\right)}dx .
\end{array}
\end{equation}

Using Eq. (7.4.19) in \cite{tab2007}, i.e., $\int_0^\infty
{\text{erfc}(\sqrt {b x} )\exp ( - ax)dx = \frac{1}{a}\left(1 -
\sqrt {\frac{b }{{b + a}}} \right)}$, the SER expression can be more
simplified as follows

\begin{equation} \label{bEr11}
\begin{array}{l}
SER =\alpha\sum\limits_{{j_1} = 1}^{L_1} \sum\limits_{{j_2} = 1}^{L_2}  \cdots \sum\limits_{{j_N} = 1}^{L_N} \left(_{j_1}^{L_1}\right)\left(_{j_2}^{L_2}\right) \cdots \left(_{j_N}^{L_N}\right)( - 1)^{( {j_1} +  \cdots  + {j_N}+N)}
\left(1 -
\sqrt {\frac{\beta }{{\beta + K}}} \right)
\end{array}
\end{equation}

In \cite{ado}, for a dual-hop relaying system with relay selection a
closed form SER expression is derived in terms of the incomplete
beta function. Here, we derived the SER of the clustered multi-hop
relay network with an arbitrary number of hops without the use of the beta function. In Section V, the SER
performance of CDFMR network is evaluated over Rayleigh fading
channels.

\subsection{Probability of SNR Gain}
In this Subsection, we present a closed-form expression for the
probability of the SNR gain achievable by the CDFMR network under consideration over the
single-hop (direct link) transmission. This performance metric is
utilized to quantify the SNR gain of relaying networks \cite{psnr}
and is defined by a ratio of the end-to-end SNR through relays to
the SNR of the single-hop transmission without a relay. The
probability of SNR gain achieved by multi-hop systems over direct
transmission could be defined as

\begin{equation} \label{eqq1}
\begin{array}{l}
\Omega  = {\rm{Pr}}\left\{ {\frac{{{\gamma _t}}}{{{\gamma _d}}} > \mu } \right\} = 1 - \int_0^\infty  {{F_{{\gamma _t}}}\left( {\mu x} \right){f_{{\gamma _d}}}\left( x \right)} dx\\
\end{array}
\end{equation}
where ${f_{{\gamma _d}}}\left( x \right) = \frac{1}{{{\Gamma
_d}}}\exp (\frac{{ - x}}{{{\Gamma _d}}})$ is the PDF of direct link
SNR with the average $\Gamma_d$, and $\mu$ is a predefined threshold
for SNR gain. Using (10), the probability of SNR gain can be
calculated as follows:
\begin{equation} \label{eqq2}
\begin{array}{l}
\Omega   = \frac{-1}{\Gamma_d} \sum\limits_{{j_1} = 1}^{L_1} \sum\limits_{{j_2} = 1}^{L_2}  \cdots \sum\limits_{{j_N} = 1}^{L_N} \left(_{j_1}^{L_1}\right)\left(_{j_2}^{L_2}\right) \cdots \left(_{j_N}^{L_N}\right)( - 1)^{( {j_1} +  \cdots  + {j_N}+N+1)} \int_0^\infty  {\exp ( - (K\mu  + \frac{1}{{{\Gamma _d}}}))} dx\\
\hspace{3.5mm}={ \sum\limits_{{j_1} = 1}^{L_1} \sum\limits_{{j_2} =
1}^{L_2}  \cdots \sum\limits_{{j_N} = 1}^{L_N}
\left(_{j_1}^{L_1}\right)\left(_{j_2}^{L_2}\right) \cdots
\left(_{j_N}^{L_N}\right)( - 1)^{( {j_1} +  \cdots  +
{j_N}+N)}}\frac{1}{K\mu\Gamma_d+1}
\end{array}
\end{equation}
Note that as we have adopted TDMA in our multi-hop system, i.e., at each time-slot only transmission in one hop is active, then the power consumptions in the multi-hop system and the one-hop system (without intermediate relays) are the same. Hence, the quantified SNR gain is indeed a measure of energy efficiency of the multi-hop network.

\section{Asymptotic performance Analysis}
In this Section, we derive simple closed-form analytical expressions
for the outage and SER performance of the CDFMR network in high SNR
regime. High SNR approximations are especially
useful for the performance analysis of data communications scenarios
in which the effect of fading is severe, and therefore a large
average SNR is necessary to achieve a target bit error rate.
Moreover, asymptotic expressions could shed light on the diversity
order of the system and may be used for resource allocation purposes
among the relay nodes.

\subsection{Asymptotic SER Behavior}
In [29], a general parametric model has been proposed to evaluate
the asymptotic behavior of the SER of communication schemes over
fading channels. In this model, it is shown that under appropriate
conditions on the PDF of the end-to-end system SNR, the behavior of
the SER, as the average SNR goes to infinity, can be characterized by
the behavior of the pdf $f_{\gamma_t}(x)$ as $x \to 0^+$. Following
a similar approach, here we derive the asymptotic SER performance of
CDFMR system in high SNR regime. From \eqref{MOD}, the SER can be
written as
\begin{equation}\label{sErjj}
\begin{array}{l}
SER = 2\alpha \int_0^\infty  {Q(\sqrt {2\beta x} ){f_{{\gamma _t}}}(x)dx}  \\
\hspace{9mm}=  2\alpha \int_0^\infty  Q(\sqrt {2\beta x} )
\sum\limits_{{j_1} = 1}^{L_1} \cdots \sum\limits_{{j_N} = 1}^{L_N}
\left(_{j_1}^{L_1}\right)\cdots \left(_{j_N}^{L_N}\right)( - 1)^{(
{j_1} +  \cdots  + {j_N}+N)} {K\exp \left( - Kx\right)}dx
\end{array}
\end{equation}
Using Maclaurin series expansion for the exponential
function \cite{tab2007}, we rewrite the SER as follows
\begin{equation}\label{sErjj}
\begin{array}{l}
SER = 2\alpha \sum\limits_{{j_1} = 1}^{L_1} \cdots
\sum\limits_{{j_N} = 1}^{L_N} \left(_{j_1}^{L_1}\right)\cdots
\left(_{j_N}^{L_N}\right)( - 1)^{( {j_1} +  \cdots  + {j_N}+N)} {
\int_0^\infty  Q(\sqrt {2\beta x} )K \left[ \sum
\limits_{l=0}^{\infty}\frac{(- Kx)^l}{l!} \right]}dx
\end{array}
\end{equation}

For $z=1,2,...,{L_m-1}$, with $L_m = \min (L_1, L_2, \cdots,
L_N)$, we have\begin{equation}\label{mosavi}
\begin{array}{*{20}{l}}
{ \sum\limits_{{j_1} = 1}^{L_1} \cdots \sum\limits_{{j_N} = 1}^{L_N}
\left(_{j_1}^{L_1}\right)\cdots \left(_{j_N}^{L_N}\right)( - 1)^{(
{j_1} +  \cdots  + {j_N})}{K^z} = 0,}
\end{array}
\end{equation}
hence, the SER expression can be more simplified as

\begin{equation}\label{sErjj}
\begin{array}{l}
SER = 2\alpha \sum\limits_{{j_1} = 1}^{L_1} \cdots
\sum\limits_{{j_N} = 1}^{L_N} \left(_{j_1}^{L_1}\right)\cdots
\left(_{j_N}^{L_N}\right)( - 1)^{( {j_1} +  \cdots  + {j_N}+N-1)} {
\int_0^\infty  Q(\sqrt {2\beta x}
)\left[\sum\limits_{l=L_m-1}^{\infty}
\frac{(-K)^{l+1}(x)^{l}}{(l)!}\right]}dx
\end{array}
\end{equation}
We next evaluate the integral term in (25). Let $X_0$ be a fixed
small positive number. The integral term in the SER expression in
\eqref{sErjj}, for any given $K$ (which is a function of indices of series) could be written as follows
\begin{equation}\label{eq24}
\begin{array}{l}
I = \int_0^\infty  Q (\sqrt {2\beta x} )\left[ {\sum\limits_{l = {L_m} - 1}^\infty  {A_l(K){x^l}} } \right]dx\\
 = \int_0^{{X_0}} Q (\sqrt {2\beta x} )\left[ {\sum\limits_{l = {L_m} - 1}^\infty  {A_l(K){x^l}} } \right]dx + \int_{{X_0}}^\infty  Q (\sqrt {2\beta x} )\left[ {\sum\limits_{l = {L_m} - 1}^\infty  {A_l(K){x^l}} } \right]dx
 \end{array}
\end{equation}
where $A_l(K)$ is a function of $K$, defined as
$A_l(K)=\frac{(-K)^{l+1}}{l!}$. Note that as $x \to {0^ + }$, the summation ${\sum\limits_{l = {L_m} -
1}^\infty {{{{A_l(K)x^l}}}{{}}} } $ can be well approximated \footnote{One could write a function $a(x)$ of $x$ as $O(x)$, if $\mathop {\lim \frac{{a(x)}}{x}}\limits_{x \to 0}  = 0$} by a
single term polynomial as $ {{{{A_l(L_m-1)x^{{L_m} - 1}}}}{{}} +
O({x^{{L_m} - 1}})}$. Using this approximation, \eqref{eq24} could
be rewritten as
\begin{equation}\label{eq25}
\begin{array}{l}
I = \int_0^{{X_0}} Q (\sqrt {2\beta x} )\left[ {A_l({L_m} - 1){x^{{L_m} - 1}} + O({x^{{L_m} - 1}})} \right]dx + \\
\int_{{X_0}}^\infty  Q (\sqrt {2\beta x} )\left[ {\sum\limits_{l = {L_m} - 1}^\infty  {A_l(K){x^l}} } \right]dx\\
 = \int_0^\infty  Q (\sqrt {2\beta x} )\left[ {A_l({L_m} - 1){x^{{L_m} - 1}} + O({x^{{L_m} - 1}})} \right]dx - \int_{{X_0}}^\infty  Q (\sqrt {2\beta x} )\left[ {A_l({L_m} - 1){x^{{L_m} - 1}} + O({x^{{L_m} - 1}})} \right]dx\\
 + \int_{{X_0}}^\infty  Q (\sqrt {2\beta x} )\left[ {\sum\limits_{l = {L_m} - 1}^\infty  {A_l(K){x^l}} } \right]dx
\end{array}
\end{equation}
Using Eq. (4a-4c) in \cite{selre}, we can see that the last two
integral terms in (27) are $O({\Gamma_m}^{-L_m})$, where $\Gamma_m =
\min (\Gamma^t_1, \Gamma^t_2, \cdots, \Gamma^t_N)$. Moreover,  using Eq.
(6.281) in \cite{tab2007} for the first integral term in (27), we have

 \begin{equation}\label{eq26}
\begin{array}{l}
 I= \int_0^\infty  Q (\sqrt {2\beta x} )\left[ {\frac{{{(-K)^{L_m}x^{{L_m} - 1}}}}{{({L_m} - 1)!}} + O({x^{{L_m} - 1}})} \right]dx=
 \frac{(-K)^{L_m}G(L_m+0.5)}{2\sqrt{\pi}(L_m!)\beta^{L_m}} +O({\Gamma_m}^{-L_m})
\end{array}
\end{equation}
where $G(\cdot)$ is the Gamma function.\\
Finally, the asymptotic SER expression could be written as
\begin{equation}\label{29}
\begin{array}{*{20}{l}}
 SER =2\alpha \sum\limits_{{j_1} = 1}^{L_1} \cdots \sum\limits_{{j_N} = 1}^{L_N} \left(_{j_1}^{L_1}\right)\cdots \left(_{j_N}^{L_N}\right)( - 1)^{( {j_1} +  \cdots  + {j_N}+N-1)} \times I\\
\hspace{9mm}= \alpha \sum\limits_{{j_1} = 1}^{L_1} \cdots
\sum\limits_{{j_N} = 1}^{L_N} \left(_{j_1}^{L_1}\right)\cdots
\left(_{j_N}^{L_N}\right)( - 1)^{( {j_1} +  \cdots  +
{j_N}+N+L_m-1)}
\frac{K^{L_m}G(L_m+0.5)}{\sqrt{\pi}(L_m!)\beta^{L_m}}
+O({\Gamma_m}^{-L_m}).
\end{array}
\end{equation}
Defining the index set $M=\{i|i\in\{1,\hdots,N\}, L_i=L_m\}$, we then have 
\begin{equation}\label{mosavi1}
\begin{array}{*{20}{l}}
SER = \alpha \left(\sum\limits_{i\in M}{\Gamma^t_i}^{-L_m}\right)
\frac{G(L_m+0.5)}{\sqrt{\pi}\beta^{L_m}} +O({\Gamma_m}^{-L_m}).
\end{array}
\end{equation}
One sees that the diversity order of this system is limited by the
minimum number of relays in clusters, i.e., $L_m$. For the special
case of system with $L_1=L_2=\cdots=L_N=L$, the SER expression
reduces to
\begin{equation}\label{sadeh0}
\begin{array}{l}
SER =\alpha \left({{\Gamma^t_1}^{-L}}+...+{{\Gamma^t_N}^{-L}}\right)
\frac{G(L+0.5)}{{\sqrt{\pi}\beta^{L}}} +O({\Gamma_m}^{-L})
\end{array}
\end{equation}
For the more special case of system with
$\Gamma_1=\Gamma_2=\cdots=\Gamma_N=\Gamma_{N+1}=\Gamma$, and
$L_1=L_2=\cdots=L_N=L$, the SER expression reduces to
\begin{equation}\label{sadeh01}
\begin{array}{l}
SER = \frac{\alpha\left(2^L+N-1\right){G(L + 0.5)}}{{\sqrt \pi {\left(\beta\Gamma\right) ^L}}}+O({\Gamma^{-L}})\\
\end{array}
\end{equation}
The dependence of SER to the number of relays in each cluster, $L$, the number of clusters, $N$, and the average SNR of each hop $\Gamma$ is evident in \eqref{sadeh01}.

\subsection{Asymptotic Outage Analysis}\label{aas}
By using a Maclaurin series expansion for the exponential function,
the expression for outage probability in Eq. \eqref{oUt} can be
rewritten as follows
\begin{equation}\label{oUt1}
\begin{array}{l}
 P_{out}  = 1 +\sum\limits_{{j_1} = 1}^{L_1} \sum\limits_{{j_2} = 1}^{L_2}  \cdots \sum\limits_{{j_N} = 1}^{L_N} \left(_{j_1}^{L_1}\right)\left(_{j_2}^{L_2}\right) \cdots \left(_{j_N}^{L_N}\right)( - 1)^{( {j_1} +  \cdots  + {j_N}+N+1)}  \left[ \sum \limits_{l=0}^{\infty}\frac{(- KA)^l}{l!} \right]

\end{array}
\end{equation}
Using \eqref{mosavi}, we then have  
\begin{equation}\label{oUt1.5}
\begin{array}{l}
 P_{out}  = 1 +\sum\limits_{{j_1} = 1}^{L_1} \sum\limits_{{j_2} = 1}^{L_2}  \cdots \sum\limits_{{j_N} = 1}^{L_N} \left(_{j_1}^{L_1}\right)\left(_{j_2}^{L_2}\right) \cdots \left(_{j_N}^{L_N}\right)( - 1)^{( {j_1} +  \cdots  + {j_N}+N+1)}  \left[ \sum \limits_{l=L_m}^{\infty}\frac{(- KA)^l}{l!} \right]
\end{array}
\end{equation}
Then, one could rewrite the outage expression in \eqref{oUt1.5} as follows
\begin{equation}\label{oUt2}
\begin{array}{l}
 P_{out}  = \sum\limits_{{j_1} = 1}^{L_1} \sum\limits_{{j_2} = 1}^{L_2}  \cdots \sum\limits_{{j_N} = 1}^{L_N} \left(_{j_1}^{L_1}\right)\left(_{j_2}^{L_2}\right) \cdots \left(_{j_N}^{L_N}\right)( - 1)^{( {j_1} +  \cdots  + {j_N}+L_m+N+1)}   K^{L_m} \frac{A^{L_m}}{L_m!}+O({\Gamma_m}^{-L_m})
\end{array}
\end{equation}
where $\Gamma_m$, and $L_m$ are as defined in the previous Subsection. 
Based on the index set $M$, \eqref{oUt2} is rewritten as follows 
\begin{equation}\label{sadeh1}
\begin{array}{l}
P_{out}  =  \left(\sum\limits_{i\in M}{\Gamma^t_i}^{-L_m}\right) \times
{A^{L_m}}+O({\Gamma_m}^{-L_m})
\end{array}
\end{equation}
One sees in \eqref{sadeh1} that the SNR of hops in which, the number of relays is more than $L_m$, do not affect the outage probability in the high SNR regime.  
For the special case of system with $L_1=L_2=\cdots=L_N=L$, the outage expression reduces to
\begin{equation}\label{sadeh2}
\begin{array}{l}
P_{out}  =  ({\Gamma^t_1}^{-L}+\cdots +{\Gamma_{N}^t}^{-L}) \times {
A^{L_m}}+O({\Gamma_m}^{-L_m})
\end{array}
\end{equation}
For the more special case of system with
$\Gamma_1=\Gamma_2=\cdots=\Gamma_N=\Gamma_{N+1}=\Gamma$, and
$L_1=L_2=\cdots=L_N=L$, the outage expression reduces to
\begin{equation}\label{ak}
{P_{out}} = \left( {{2^{L}} + N - 1} \right) \times {\left(
{\frac{A}{\Gamma }} \right)^{L}}+O({\Gamma}^{-L}) .
\end{equation}
In Section \ref{SiM}, the performance of CDFMR network with ad-hoc routing is evaluated using numerical and simulation results.
\begin{figure}[h]
\centering
\includegraphics[width=3.5in]{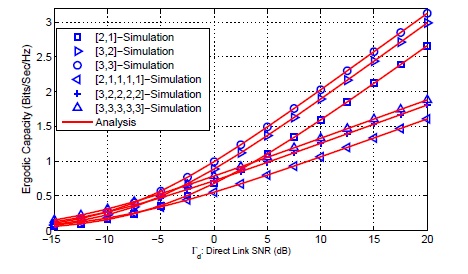}
\caption{Ergodic capacity versus the source-destination average SNR, $\Gamma_d$. }
\label{ergo}
\end{figure}

\begin{figure}[h]
\centering
\includegraphics[width=3.5in]{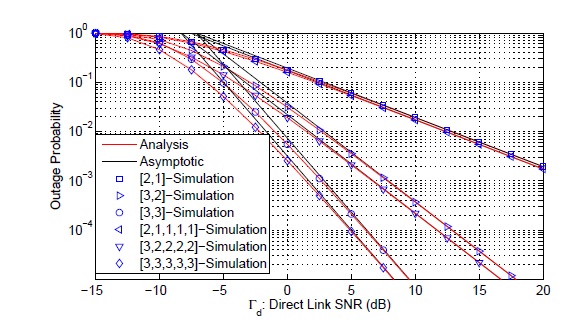}
\caption{Outage probability versus the source-destination average SNR, $\Gamma_d$.}
\label{outa}
\end{figure}

\begin{figure}[h]
\centering
\includegraphics[width=3.5in]{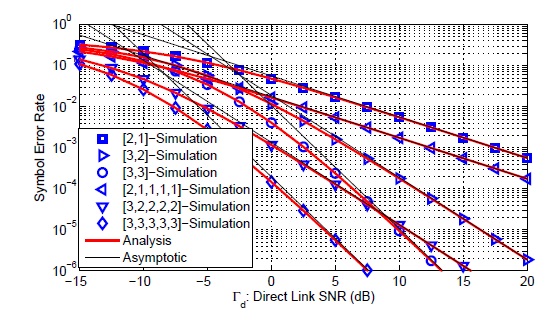}
\caption{SER performance of CDFMR network with BPSK modulation as a
function of the source-destination average SNR, $\Gamma_d$.} \label{ser}
\end{figure}

\begin{figure}[h]
\centering
\includegraphics[width=3.5in]{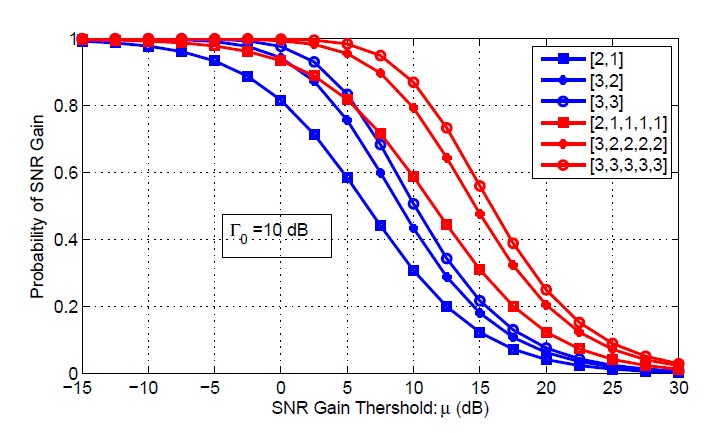}
\caption{Probability of SNR gain for CDFMR network with balanced links (BL), and unbalanced links (UBL) as a function of the SNR gain threshold, $\mu$.} \label{psnr}
\end{figure}

\section{Performance Evaluation}\label{SiM}
In this Section, we investigate the performance of the CDFMR network with ad-hoc routing
in terms of probability of outage, ergodic capacity, SER and
probability of SNR gain based on the derived analytical results and
simulations. The first two metrics quantify how efficiently the spectrum is utilized and the last two are indicators of energy efficiency of the multi-hop network. We describe the structure of the system model with
the notation $[L_1,L_2,...L_{N}]$, where $N$ is the number of relay 
clusters, and $L_i$ is the number of relay nodes in $i$th cluster, as described
in Section II. In this Section, we evaluate the performance of several different CDFMR systems
including the $3$-hop networks $ [2,1], [3,2], [3,3],$ and $6$-hop networks $[2,1,1,1,1] , [3,2,2,2,2],$ and $[3,3,3,3,3]$. For modeling unbalanced links in clustered multi-hop system, we adopt the method of \cite{gfa}, and assume that the $k$th terminal, $k$ = 1,..., $N+1$ is
located at the distance $d_k=\frac{2k}{(N+1)(N+2)}d$ from its
previous terminal, where $d$ is the distance between source and destination. Hence, using the Friis propagation formula
\cite{rap}, the average SNR of $k$th hop is given by
$\Gamma_k=\left(\frac{(N+1)(N+2)}{2k}\right)^{\delta}\Gamma_d$,
where $\delta$ is the path loss exponent and $\Gamma_d$ is the average SNR of direct link.\\
Fig. \ref{ergo} shows the ergodic capacity of the CDFMR network as a
function of the average SNR per hop for 3-hop, and 6-hop systems,
and $\delta=4$. As evident, the analytical results based on Eq.
\eqref{eRg} are consistent with simulation results. From this figure
one sees that for a fixed $L_m$ (minimum number of relays in
clusters), as the number of hops (or relay clusters) increases, the
ergodic capacity decreases $(N=5$ vs. $N=2)$. Note that by
adding more relay clusters between source and destination, when the
source-destination distance is constant, the factor $K$ in
\eqref{eRg} assumes larger values, but because of the factor
$\frac{1}{N+1}$  for the half-duplex operation of relays,
the ergodic capacity decreases. Another observation is that for fixed number of hops,
$N+1$, as the minimum number of relays in clusters (diversity
paths$= L_m$) increases, the ergodic capacity also increases.

Fig. \ref{outa} shows the outage probability as a function of the
average SNR per hop, $\gamma$, for 3-hop and 6-hop systems with
the predefined outage rate $R=$0.3 bit/s/Hz. This figure
indicates that the analytical results based on Eq. \eqref{oUt}
closely follow the simulation results. In simulations, we first
generate Rayleigh distributed channel gains for the CDFMR network
and then compare the instantaneous data rate of the system with the
outage rate. As evident, with constant source-destination distance,
 when the number of relay clusters, $N$, increases, the outage probability improves; however,
adding more relay clusters does not affect the slope of the outage
probability curves. On the other hand, a larger minimum number of
relays in clusters, $L_m$, enhances the slope of the outage curves
and hence significantly improves the performance. This is because 
the diversity order of system is limited by the minimum number of relays in clusters. 
In Fig. \ref{outa}, the dashed lines indicate the asymptotic outage probability in high
SNR regime based on Eq. \eqref{sadeh1}, which precisely match both analytical and simulation results.

Fig. \ref{ser} depicts the SER performance of the CDFMR network with
ad-hoc routing and BPSK modulation as a function of the average SNR per hop. As
evident, the numerical results based on the analysis in Section
III.B perfectly match with the simulation results. We also observe
that (i) for the same $L_m$, increasing the number of hops, $N$,
results in a lower system SER; and (ii) for the same $N$, increasing
the minimum number of relays in clusters decreases the SER with a
rate proportional with the diversity order. In Fig. \ref{ser}, the asymptotic
SER performance based on Eq. \eqref{mosavi1} are also depicted using dashed lines.
One could see that in high SNR regime, the proposed asymptotic SER expression perfectly follows the analysis presented in Section III.D. \\

The probability of SNR gain for multi-hop system over direct
transmission in a CDFMR network with ad-hoc routing is depicted for balanced and unbalanced links scenarios in Fig. \ref{psnr}. As evident, in comparing the probabilities of SNR gain for different communication scenarios the benefit of multi-hop systems (energy efficiency) increases with (i) more
hops between source and destination nodes, and (ii) more relays in
each cluster. In the balanced and unbalanced link scenarios under consideration, the longest hop distance is respectively $\frac{1}{N+1}$ or $\frac{2(N+1)}{(N+1)(N+2)}=\frac{2}{N+2}$ for $d=1$. This translates to weakest average SNR per hop of $\Gamma_k=(N+1)^\delta\Gamma_d$, and $\Gamma_{N+1}=\left(\frac{N+2}{2}\right)^{\delta}\Gamma_d$. The larger value of the former clearly explains the superior performance of the CDFMR network with balanced links in Fig. \ref{psnr}, as the weakest hop dominates the performance in this decode-and-forward wireless network. 

Examining Figs. \ref{ergo}-\ref{psnr}, for a given number of relays per cluster, it is expectedly evident that increasing the number of relay clusters (hops) between a given source and destination pair, improves the reliability of communications at the cost of reduced average rate. In other words, increasing the number of hops reduces both the average and the variations of the capacity of the CDFMR network with ad-hoc routing, but improves the reliability or accordingly the power (energy) efficiency of the network. The former manifests itself as a smaller pre-log factor for a greater number of hops in ergodic capacity.

\section{Conclusions}\label{ConclusionSection}
In this paper, the performance of clustered decode-and-forward
multi-hop transmission networks with ad-hoc routing is analyzed, and the effects of different design parameters on network energy and spectrum efficiencies are investigated. We consider a general heterogeneous network topology where the relay clusters are arbitrarily distanced between source and destination and may contain different numbers of nodes.  In
particular, closed form analytical expressions for the PDF of the
end-to-end system SNR, the ergodic capacity, the outage probability, the symbol error
rate, and the probability of SNR gain have been derived. The
provided analyses are exact and enable the performance evaluation in
all low, medium, and high SNR regimes. Moreover, simple asymptotic
expressions for the outage probability and SER performance in high
SNR regime have been presented. The results reveal that increasing
the number of hops or the minimum number of relays in intermediate
clusters reduces the outage probability and improves the SER performance of
the system. In addition, the proposed analysis quantifies how ergodic capacity and the probability of
SNR (power) gain improve with increasing the minimum number of relays in
clusters. 

\ifCLASSOPTIONcaptionsoff
\newpage
\fi
\bibliographystyle{wileyj}
\bibliography{IEEEabrv,bibl}
\end{document}